\documentstyle[12pt,a4]{article}
\nopagebreak

\def \ftp#1#2#3{\Psi_{#1}(\vec #2,\vec #3)}
\def \diff#1{\Bigl({\partial \over {\partial z}} - ik_#1 \Bigr)}

\begin{document}


\centerline{\Large\bf{Colour transparency in hadronic basis}}
\vskip1.5cm
\centerline{\large{A. Bianconi, S. Boffi and D.E. Kharzeev}
\footnote{On leave from Moscow University, Moscow, Russian Federation}}
\vskip0.5cm
\hskip0.9cm\small{Dipartimento di Fisica Nucleare e Teorica, Universit\`a di
Pavia, and}

\hskip0.9cm\small{Istituto Nazionale di Fisica Nucleare,
Sezione di Pavia, Pavia, Italy}
\vskip3.5cm
\begin{abstract}
The dynamics of nuclear transparency in
hard nuclear reactions is studied by an expansion of
the correlator of the hard scattering operator on a hadronic basis.
Colour transparency appears as an effect of interference between
the amplitudes corresponding to the final state interaction of
different intermediate baryonic states. We find that colour
transparency occurs
even at moderate momentum transfers  if
one chooses  the missing momentum of the struck nucleon
by appropriate kinematics.
 In general, the role of the Fermi-motion is found to be crucial;
without it, any effect of anomalous nuclear transparency would be impossible.
\end{abstract}
\clearpage

The concept of colour transparency [1-5] has attracted great interest in the
past few years (see, for example, [6-10]). This concept is based essentially on
two ideas:

$i$) high-momentum transfer nuclear reactions select from the hadron wave
function the ``compact'' component, whose size is smaller than the average
size of the hadron;

$ii$) since pointlike colourless states do not interact in QCD, this compact
state will undergo small initial- and final-state interactions with the
nuclear medium.

The property of colour transparency is a direct consequence of QCD
as a gauge theory and it should manifest itself at high enough energies and
momentum transfers. However, very important things are still unclear.

Which energy is ``high enough'' to enter into the colour transparency domain?
How the transition between regimes of normal nuclear attenuation and colour
transparency occurs?
Simple answers to these questions are still lacking.

Indeed, the only experiment [11] dedicated to the study of colour transparency
in $(p,2p)$ reaction, has demonstrated an unexpected oscillatory pattern
of the ratio of nuclear to hydrogen cross section. This is in a
striking contradiction with theoretical expectations, based on the
naive picture of a compact state expanding in the nuclear medium after
the hard interaction has occured. Several attempts to find out a reason of the
oscillations inside of the mechanism of hard $pp$ scattering were made
[7,12-14]. However, it is yet unclear whether or not colour transparency should
manifest itself at all in the kinematical conditions of experiment [11].
The more detailed, dynamical treatment of the process is thus needed.

To get an insight into the problem, let us note firstly that all the essential
information about
the space-time evolution of a hadronic state produced by the hard
process is contained in the (euclidian: $\tau^2=-x^2$) correlator of
the hard scattering operator $\hat J$ inside a proton:
$$\Pi_{J}(\tau)=<p|T\{\hat J(x)\hat J(0)\}|p>.  \eqno(1)$$

This correlator can be represented as a sum of contributions arising from
different baryonic states. At small distances the correlator
is composed by a great number of baryonic excitations. On the other hand,
in the large distance limit the behaviour of the correlator $\sim \exp(-m\tau)$
is governed by the mass $m$ of the ground baryonic state (proton).
The proper formation time $\tau_{form}$ of the proton thus can be defined
as a scale at which the correlator
approaches this asymptotic behaviour [15] and can be roughly estimated as
$\tau_{form} \sim$ $(m_{*}-m)^{-1}$, where $m_{*}$ is the mass of the lowest
baryonic excitation with proton quantum numbers.
The colour transparency can manifest itself if the baryonic state
produced by hard scattering interacts with the nuclear medium before
it has turned into the proton.

The effect of nuclear medium on the correlator (1) is to attenuate each
baryonic state component as well as to cause transitions between them.
As we shall demonstrate, this latter effect is the one that
ensures the colour transparency property in hadronic basis.

The final state
interaction with nuclear medium $\hat V$ is soft in the sense that the
time scale $\tau_{int}$ at which the correlator
$$\Pi_{V}(\tau)=<p|T\{\hat V(x)\hat V(0)\}|p>   \eqno(2)$$
reaches its asymptotics is large: $\tau_{int} > \tau_{form}$.
This is true for all the phenomenological mean field nuclear potentials,
since their depth $V$ does not exceed the mass spacing between the proton
and the first baryonic levels: $V < (m_{*}-m) \sim$ 300 MeV.
Therefore, only transitions between the proton and the first baryonic
excitation in the intermediate state are
effective in the nuclear medium. In this case a simple and economic
description can be made in hadronic basis. The dynamics of the
process can therefore be described by a system of two coupled equations.

To put this idea on a quantitative basis, we restrict ourselves to the
consideration of two baryonic states, namely the proton and its
excitation with the same quantum numbers. The reason for this
is the following. In vacuum only baryonic
states with the same quantum numbers can mix in the correlator (1).
The nuclear medium can, in general, cause transitions between
baryonic states with different quantum numbers; however it is well known
(see, for example, reviews [16,17]) that at high energies diffractive hadronic
interactions are dominated by the Pomeron
scalar exchange which again favours transitions between states with the
same quantum numbers. In practical evaluations we have used
the mass of the lowest state
with proton quantum numbers, i.e. of the Roper resonance $N(1440)$.

Probably the most interesting feature of the approach described above
is that it helps to understand the
fundamental importance of Fermi-motion for colour transparency.

Let us illustrate this point in detail. Consider a hypothetical nucleus
where nucleons have fixed zero three-momentum, i.e. no Fermi-motion exists.
The four-momentum of nucleons in the lab is therefore simply $p_0=(m_N,0)$.
Suppose also that the  kinematics of the process allows one
to fix the transferred momentum $q$.
In this case the invariant mass of the state produced by hard scattering is
strictly determined:
$ s=s_0=(p_0+q)^2 $. Substituting this constant value of $s_0$
into the correlator (1) one finds that it is composed by only one
hadronic state of mass $m_0=\sqrt s_0$ (if it exists). This state
obviously will be attenuated in the nuclear medium in a usual way since its
wave
function is of the normal hadronic size: colour transparency cannot
appear.

The existence of Fermi-motion smears the distribution in the nucleon momentum
and therefore
causes an intrinsic uncertainty in the invariant mass of the produced state.
Let us consider a hard scattering producing
 a baryonic state $\mid k>$ at fixed transferred
momentum $q$ from a nuclear bound proton. The square modulus of
the invariant amplitude, summed over the residual nucleus
final states, reads

$$ \mid<k\mid\hat J(q^2)\mid i>\mid^2 =
Z \int d^3 k\  \vert g_k(q^2)\vert^2 \frac{m_k \Gamma_k}
{\left[m_k^2-(k+q)^2\right]^2+m_k^2\Gamma_k^2}\ n(k), \eqno(3)$$
where $n(k)$ is the momentum distribution of protons in the nucleus, defined
as an integral of the nuclear spectral function $S(k,E_f)$ over the residual
nucleus excitation energies
$$ n(k)=\int dE_f S(k,E_f); \eqno(4)$$
Z is the number of protons in the nucleus, and $g_k(q^2)$ is the
elementary amplitude of $\mid k>$ production on hydrogen.
It is clear from eq. (3) that in the presence of Fermi-motion the
invariant mass of the state produced by hard scattering is not strictly fixed
even at fixed momentum transfer. Owing to the Fermi-motion, at each value of
$q$ several terms in the spectral density contribute to the correlator (1).

However, at low $q$ the Fermi-motion is still unable to allow for a
composition of a sufficiently coherent wave packet. At large momentum
transfers the situation changes; the uncertainty in the invariant
mass of the produced state becomes larger than a characteristic mass spacing
between physical hadronic states, and a coherent compact wave packet
can be produced.

The excited components of the wave packet will then undergo
multiple diffractive transitions in the nucleus, accompanied by the
longitudinal momentum transfer

$$q_L \simeq {m_k^2 - m^2\over 2p} .\eqno(5)$$
The cancellation of the corresponding amplitudes with the
amplitude of proton elastic rescattering inside the nucleus
will cause colour transparency.

We have thus come to the conclusion that the Fermi-motion plays the fundamental
role in exclusive knock-out reactions, allowing for the very existence of
colour transparency.

Our treatment, as we shall demonstrate, leads to some new results.
In particular, we clarify the fundamental role of Fermi-motion in colour
transparency.\footnote{While this paper was in preparation,
we became aware of a preprint by
Jennings and Kopeliovich [18], which appeared simultaneously with our previous
work [19]. In spite of the difference in formalisms we use, the similarity
of conclusions is amazing.}

\def \ftp#1#2#3{\Psi_{#1}(\vec #2,\vec #3)}
\def \diff#1{\Bigl({\partial \over {\partial z}} - ik_#1 \Bigr)}

The dynamics of the final state can be described, in the eikonal
approximation, by the system of coupled equations:

$${\Bigl({\partial \over {\partial z}} - ik_n \Bigr)}
\Psi_n
\ =\ {1 \over {2ik_n}} \sum_{l=1,2}
V_{nl}\Psi_l ,\ \ n=1,2\ ,
\eqno(6)$$
$$\vec r\ \equiv\ z {\hat n}\ +\ \vec b. $$
The wave functions $\Psi_1$ and $\Psi_2$ are
the proton and the $N^*$ states
with the same energy and direction of
motion $\hat n$, but different mass and
absolute values of momentum:
$$E\ = \sqrt{m^2_*+k^2_2}\ =\ \sqrt{m^2+k^2_1},\ \ \ k_1\equiv p.
\eqno(7)$$
The effective optical potential
$\{V_{nm}\}$ describes their interactions with the
residual (A--1) nucleus.
It induces transitions
between the two nucleon states ($1 \rightarrow 2$ and $2 \rightarrow 1$)
as well as elastic interactions
(we neglect here spontaneous decay of $N^*$ inside the nucleus, which is
justified at high energies).
The momentum transfer in the $ 2 \rightarrow 1$ process is
$$q_2\ \equiv\ k_1-k_2 \ =
\ p-\sqrt{p^2-(m_*^2 -m^2)},\eqno(8)$$
with necessarily  $p^2\ >\ m_*^2 -m^2$.
We define
$\Psi_{n}(\vec r,\vec {r_o})\ (n=1,2)$ as the
hadronic state produced by the hard interaction with the momentum
transfer $\vec q$
at the point $\vec {r_o}$.
We require
$\Psi_n(\vec {r_o},\vec {r_o})\ =\ B_n\  (n=1,2)$,
with $|B_1|^2+|B_2|^2$ = 1. The outgoing elastic-channel wavefunction
$\Psi_{\vec p}(\vec r)$ is the
coherent sum of all the contributions from every $\vec {r_o}$
inside the nucleus, with relative phase given by the exponent
$e^{i\vec q\cdot\vec {r_o}}$, weighted
by the bound state function $ \phi_o(\vec {r_o})$.
For $z\ >>\ R$ (nuclear radius) eqs. (6) imply
$$\Psi_1(\vec r,\vec {r_o})\
\equiv\
\Psi[(z,\vec b),(z_o,\vec {b_o})]\
\simeq f(\vec b,\vec {b_o},z_o) e^{ipz}. \eqno(9)$$
Then the cross section for the process will be proportional
to $|I|^2$, where
$$I(z)\ =
\ \int d^3 r_o e^{i \vec q\cdot \vec {r_o}}
\phi_o(\vec {r_o}) \Psi_1[(z,\vec {b_o}),(z_o,\vec {b_o})]
\eqno(10)$$
and $|I|^2$ is independent of $z$ for $z$ larger
than the nuclear potential range.

The shape of the functions $V_{ij}$ is a critical point. At high energies for
$V_{11}$ we have the Glauber expression
$${{V_{11}(z,\vec b)}\over{2ik_1}}\ =
\ -{{2\pi iA}\over{k_1}}f_{11}(0)\rho(z,\vec b)\
\simeq\ - {A \over 2}\sigma_{tot}\rho(z,\vec b),
\eqno(11)$$
where $A$ is the nuclear mass number,
$\rho$ is the nuclear density normalized to 1,
and $f_{11}(0)$ is the elastic forward transition
amplitude; for $\sigma_{tot}$ we have used the value $\sigma_{tot}\ =\ 40$ mb.
Similarly $V_{22}/V_{11}$ =
$\sigma^{tot}_2/\sigma^{tot}_1$. Since $\sigma^{tot}_2$ is
unknown, we assume $V_{22}$ = $V_{11}$.

For $V_{12}$ a first guess could be
$V_{12}(z,\vec b)$ $\equiv$
$V_{11}(z,\vec b) f_{12}(0)/f_{11}(0)$.
In homogeneous nuclear matter this potential
would be $z$-independent, allowing for
solutions of the kind $\hbox{\rm const} \cdot \exp[(ik-b)z]$.
This does not agree with the change of momentum
$k_2\rightarrow k_1$ during rescattering.
Neglecting deflection and assuming
axial symmetry we require
$$<1,k_1|V(b)|2,k_2>\ =\ {{f_{12}(0)}\over {f_{11}(0)}}
<1,k_1|V(b)|1,k_1>.\eqno(12)$$
Thus we obtain

$$<1,z|V(b)|2,z'>\ \equiv V_{12}(z,b)\delta(z-z')=
\ e^{i(k_1-k_2)z} V_{11}(z,b)\delta(z-z')
{{f_{12}(0)}\over {f_{11}(0)}}.\eqno(13)$$

The coefficients $B_n$ depend in
a critical way on the detailed kinematics of the hard scattering.
As far as their modulus is concerned we can write the
general expression as

$$|B_i|^2\ =\ \int d^3 k  \vert g_i(t)\vert^2 R_i(s) n(k). \eqno(14)$$
In eq. (14) the variable $s\ \equiv\ (q+k)^\mu(q+k)_\mu$
is the squared mass of the intermediate state particle, and
$R_i(s)$ is the
weight for producing the baryonic state $i$ (not far from the
resonance peak),

$$R_i(s)\ =\ {{m_i\Gamma_i} \over {(s-m_i^2)+m_i^2\Gamma_i^2}}.
\eqno(15)$$
In particular for the elastic channel $\Gamma \rightarrow 0$
and $R_p(s)=\pi\delta(s-m_p^2)$. The function
$g_i(t)$ is a vertex coupling dependent on the momentum
transfer $t$.
At high $t$ the quark counting rules [20,21] state, on
a general ground, that this factor should only depend
on the number of valence quarks participating in the interaction.
For every baryon ground or
excited state we have the same coupling $g(t)$.
At lower energies the same result can been obtained
within the nonrelativistic quark model [8].
Since only the ratio $B_2/B_1$ is important here we can assume
$$\vert g_1(t)\vert^2 \ =\ \vert g_2(t)\vert^2 .\eqno(16)$$

The quantities $B_1$ and $B_2$ defined above are functions of the
missing momentum
$$\vec p_m\ \equiv\ \vec p\ -\ \vec q .\eqno(17)$$

In the special case of parallel kinematics:
$\vec p$ $\parallel$ $\vec q$. Then $p_m = p-q$ and

$$s - m_1^2\
=\ 2q (p_m - k_z) + p_m^2 -k^2+m_p^2-m_i^2\ $$
$$\simeq\ 2q ({p_m} -k_z)+m_p^2-m_i^2 . \eqno(18)$$
In numerical calculations we use the exact formula,
but the last approximate expression inserted in eqs. (15) and (14)
makes it clearer the role of events with given missing momentum
in exclusive $(e,e'p)$ experiments:
$$|B_i|^2\ \propto\ \int d^3 k \ n(k)
{{m_i\Gamma_i} \over {[2q(p_m-k_z)+m_p^2-m_i^2]^2+m_i^2\Gamma_i^2}}.
\eqno(19)$$
Incidentally we note that at the leading order
in $p_m/q$, $|B_i|^2\ =\ |B_i(p_m,q)|^2\ =\ |B_i(p-q,q)|^2$
only depends on $\vec p_m$ via its component
along $\vec q$, $(p_m)_z$.
The values of $\vert B_2\vert^2\equiv \vert B_*\vert^2$
for $q = 4, 8, 40$ GeV are depicted as functions of $p_m$ in fig. 1.
The parameters chosen for the Roper resonance are
$m_*$ = 1.44 GeV, $\Gamma_*$ = 350 MeV.

The integral in eq. (19) attains its
maximum when $k_z\equiv 0$, i.e.  $p_m = (m_*^2-m_p^2)/2q > 0$.
This behaviour is particularly remarkable for low $q$.
Of course,  for $q \rightarrow \infty$ we have
$(m_*^2-m_p^2)/2q$ $\rightarrow$ 0 and the
$|B_*|^2$ distribution tends to look like the $|B_p|^2$
distribution, symmetric around $p_m = 0$. This
has the strongest consequence on transparency.

We have
assumed that the $B$ coefficients have the same phase,
which can be strictly justified only in the case of $ep$
interactions and neglecting the finite width of the resonance.
This last assumption is not crucial, however.

Numerical calculations have been performed for the
$^{27}$Al target nucleus. Then one has to consider bound
states up to $l=2$. To represent them we
use harmonic oscillator wavefunctions.
The matter density in eq. (11) has been
assumed to be an isotropic Wood-Saxon function with
radius 1.1$\cdot A^{1/3}$ fm and diffuseness 0.5 fm.

The obtained values of transparency are plotted in fig. 2
versus $p_m$ (in units of the Fermi momentum $p_f$)
 for $q = 4, 8$ and
40 GeV, and in fig. 3 versus $q$  for
$p_m/p_f = 0,\  0.35 p_f$ and 0.7$p_f$.

Figs. 2 and 3 can be related to fig. 1 where  the probability
$|B_*|^2$ of the $N_*$ excitation
is plotted for the same values of $m_*$ and $q$
and in the same $p_m$ range of fig. 3.

It is easy to see that the maxima of transparency
in fig. 2 occur at practically the same values of $p_m$
as the maxima of the corresponding curves in fig. 1.
So an obvious conclusion is that maxima of transparency
are attained for those values of $q$ and $p_m$ where
$|B_*|^2$ is maximum.

The non trivial conclusion is that only for $p_m \sim 0$
transparency is attained at asymptotic momentum transfer.
At non-asymptotic $q$ transparency maxima
manifest themselves at other (positive) values of missing
momenta. These maxima  are strictly
related to the maxima of the probability $|B_*|^2$ of
exciting baryonic resonance in the hard scattering vertex.

\vskip 0.5cm
\centerline{\Large{\bf Figure captions}}
\vskip 0.5cm

Fig. 1. The values of the squared amplitude $\vert B_*\vert^2$ of
producing a Roper resonance as a function of the missing momentum $p_m$ in
units of the Fermi momentum ($p_f = 234$ MeV). Dotted, dashed and
solid curves for $q = 4, 8, 40$ GeV, respectively.

Fig. 2. The transparency coefficient as a function of missing momentum
$p_m$ in units of the Fermi momentum ($p_f = 234$ MeV).
Dotted, dashed and solid curves for $q = 4, 8, 40$ GeV, respectively.

Fig. 3. The transparency coefficient as a function of the momentum
transfer $q$. Dotted, dashed and
solid curves for $p_m/p_f = 0.0, \ 0.35, \ 0.7$.

N.B. Figures available from the authors

\end{document}